\documentclass[11pt]{asaproc}

\usepackage{graphicx}
\usepackage{amsmath, amssymb}
\usepackage{natbib}

\usepackage{tikz} 
\usepackage{xcolor,color}
\usetikzlibrary{fit}

\usepackage[colorlinks=true,
			citecolor=blue,
			linkcolor=blue,
 			urlcolor=blue,
            breaklinks=true]{hyperref} 

\usepackage{times}

\title{Estimating Propensities of Selection for Big Datasets via Data Integration}

\author{Lyndon Ang\thanks{Australian National University, Canberra, ACT, 2600, Australia} \and  Robert Clark\footnotemark[1] \and  Bronwyn Loong\footnotemark[1] \and Anders Holmberg\thanks{Australian Bureau of Statistics, Belconnen, ACT, 2617, Australia}}
\begin{document}

\maketitle

\begin{abstract}
    Big data presents potential but unresolved value as a source for analysis and inference.  However, selection bias, present in many of these datasets, needs to be accounted for so that appropriate inferences can be made on the target population.  One way of approaching the selection bias issue is to first estimate the propensity of inclusion in the big dataset for each member of the big dataset, and then to apply these propensities in an inverse probability weighting approach to produce population estimates.  In this paper, we provide details of a new variant of existing propensity score estimation methods that takes advantage of the ability to integrate the big data with a probability sample.  We compare the ability of this method to produce efficient inferences for the target population with several alternative methods through a simulation study.

\begin{keywords}
Big Data, Selection Bias, Propensity Score Estimation, Inverse Probability Weighting
\end{keywords}

\end{abstract}

\section{Background and Motivation}\label{intro}

As a byproduct of the ongoing digitalisation of society, there is an increasing amount of data being collected about individuals, entities and the environment.  These new data sources include ``big data'' from sensors, satellites, administrative systems, and so on.  Statistical agencies have started exploring how to harness these data sources for producing statistics.  For example, the Australian Bureau of Statistics (ABS) has in recent years utilised employee payroll data submitted by employers to the Australian Taxation Office (ATO) to help produce wages and jobs statistics (\citeauthor{australianbureauofstatistics_2024}, \citeyear{australianbureauofstatistics_2024}, \citeyear{australianbureauofstatistics_2024c}), bank transactions data to assist in producing statistics on household spending (\citealt{australianbureauofstatistics_2024a}), and agricultural levies data to produce statistics on agricultural production (\citealt{australianbureauofstatistics_2024b}).

Big data is generated primarily for non-statistical purposes.  As such, these datasets may not accurately represent the population of interest and are more akin to non-probability samples (\citealt{golini.righi_2024}).  Specifically, they may suffer from selection bias where some groups in the target population are less likely to be included in the dataset.  Although they may be large, when these datasets are used directly without adjustment to estimate population parameters such as totals and means, biased estimates are likely to be produced (see for example \citealt{bethlehem_2010} and \citealt{meng_2018}).

Various approaches have been developed in recent years to address selection bias in non-probability data; see \citet{zhang_2019}, \citet{wu_2022a}, and \citet{yang.kim_2020a} for some reviews on developments in this area.  One class of methods are the so-called quasi-randomisation approaches, which assume that the missingness in the big data is governed by an unknown but discoverable selection mechanism.  The idea is to use available auxiliary information from the population (see for example, \citealt{burakauskaite.ciginas_2023}) or from a supplementary \textit{reference} probability sample (see for example, \citealt{chen.etal_2020} and \citealt{wang.etal_2021}) to estimate propensities of selection into the big data.  The inverse of these propensities are then used as weights to be applied in a weighting process for the big data to estimate finite population quantities.

In some scenarios, it may be possible to identify which units in the reference sample are also present on the big dataset.  This would occur, for example, if a common unit identifier exists and is available on both the big dataset and the reference sample.  In the establishment data setting, for example, a unique business number often exists for administrative purposes such as the filing of taxation data.  In this paper, we present an approach to estimating the propensities of selection for a non-probability sample (and in particular a big dataset) which takes advantage of the ability to link records on the reference sample to the big dataset.  We compare the efficiency of the method to a number of other methods in a simulation setting.

The rest of the paper is structured as follows.  We outline the basic setup for the paper in Section \ref{setup}.  In Section \ref{propscores} we provide a brief overview of some of the existing approaches to estimating propensity scores in a non-probability dataset.  Section \ref{oe_propscores} introduces our proposed propensity score estimator, while in Section \ref{simstudy} we examine the performance of the estimator in a simulation study.  Section \ref{conclusion} provides some concluding remarks.

\section{Basic Setup} \label{setup}

Let $\mathcal{U}$ be a finite population of size $N$.  For each unit $i$ in the population we have values $(\boldsymbol{x}_i,y_i)^T$ for a data item of interest $y$ and some auxiliary data items $\boldsymbol{x}$.  In this paper, we are interested in estimating the population mean $\bar{Y} = \mu_y = N^{-1}\sum_{i=1}^{N}y_i$.

Suppose we have a probability sample $A$ (which we will call the ``reference'' sample) of size $n_A$ drawn from the population.  $A$ contains information for $\boldsymbol{x}$ but not $y$.  Define $\pi^A_i=P(i \in A|\mathcal{U})$ to be the inclusion probability for unit $i$ being in the probability sample, and $d^A_i=1/\pi^A_i$ as the survey design weight for $i \in A$.  The $\pi^A_i$ and $d^A_i$ values are known from the probability sample design.

We also have a non-probability sample $B$.  $B$ contains information for $\boldsymbol{x}$ as well as $y$.  Let $\delta_i=I(i \in B)$ be an indicator variable for unit $i$ being included in $B$.  The non-probability sample size is $N_B=\sum_{i=1}^{N}{\delta_i}$.  In contrast to $\pi^A_i$, the inclusion probabilities $\pi^B_i=P(\delta_i=1|\mathcal{U})$ are unknown and need to be estimated.  Define $C=\mathcal{U} \setminus B$ to be the units in the population not included in $B$.  There may or may not be overlap between the two samples $A$ and $B$.  Figure \ref{fig:population domains} depicts these domains within $\mathcal{U}$.

\tikzset{filled/.style={fill=circle area, draw=circle edge, thick},
    outline/.style={draw=circle edge, thick}}

\setlength{\parskip}{5mm}

\begin{figure}[!h]
	\centering

	\colorlet{circle edge}{black!50}
	\colorlet{circle area}{black!20}

	\tikzset{
  		filled/.style={fill=none, draw=circle edge, thick},
  		outline/.style={draw=circle edge, thick}
	}

	\begin{tikzpicture}
		\draw[filled] (0,0) circle[radius=1.2cm] node {$A \setminus B$}
					  (0:2cm) circle[radius=1.2cm] node {$B \setminus A$};
		\node[anchor=north] at (current bounding box.south) {$A \cap B$};
		\draw[thick, ->] (1,-1.3) -- (1,-0.2);
		
		\node [draw,fit=(current bounding box),inner sep=6mm] (frame) {}; 
		\node [below left] at (frame.north west) {$\mathcal{U}$};
		\node[above right] at (frame.south west) {$C \setminus A$};

	\end{tikzpicture}
	\caption{Domains Within the Population $\mathcal{U}$}
	\label{fig:population domains}
\end{figure}
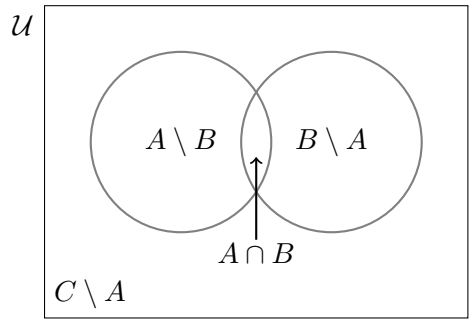

We make some assumptions regarding the selection mechanism for units being included in $B$, in order to facilitate forming inferences using those datasets; see for example \citet{chen.etal_2020} and \citet{yang.etal_2021}.  These assumptions are:

\begin{enumerate}
	\item[]
    \begin{enumerate}
        \item \textit{Ignorability}: Conditional on the set of covariates $\boldsymbol{x}_i$, $\delta_i$ and $y_i$ are independent.
        \item \textit{Positivity}:  Conditional on $\boldsymbol{x}_i$, $P(\delta_i=1|\boldsymbol{x}_i) > 0, \forall i \in \mathcal{U}$.
        \item \textit{Independence}: Conditional on $\boldsymbol{x}_i$ and $\boldsymbol{x}_j$, $\delta_i \perp \delta_j, i \neq j, \forall i,j \in \mathcal{U}$.
    \end{enumerate}
\end{enumerate}

Ignorability implies that $P(\delta_i=1|\boldsymbol{x},y)=P(\delta_i=1|\boldsymbol{x})$.  In other words, selection into $B$ is ignorable conditional on the covariates $\boldsymbol{x}$.  This assumption is similar to the Missing-At-Random (MAR) scenario of \citet{rubin_1976}.  \citet{andridge.etal_2019} and \citet{boonstra.etal_2021} refer to this type of selection process as Selection At Random (SAR).

The positivity assumption states that conditional on $\boldsymbol{x}$ every unit in the population has a non-zero chance of inclusion in $B$.  This may not always hold - for example, in a big dataset generated by a social media platform, only those persons who are members of the platform have a chance of inclusion in the big dataset.

In addition to the above, we assume that it is possible to accurately identify which units in $A$ are also present in $B$.  This assumption may be satisfied if there exists a common unit identifier available on $B$ and the reference sample $A$, or if it is possible to accurately link records in $A$ to $B$ using a common set of linking variables.  Alternatively, this assumption could be designed for by having a question in the survey instrument for $A$ which asks whether the respondent is also on the dataset $B$.

\section{Estimation of Propensity Scores}\label{propscores}

We wish to estimate the unknown $\pi^B_i$.  Suppose these propensities can be modelled parametrically so that $\pi^B_i = \pi(\boldsymbol{x}_i,\boldsymbol{\theta}_0)$, where $\boldsymbol{\theta}_0$ are the true values of the unknown model parameters.  The population likelihood function of $\pi^B_i$ is

\begin{equation} \label{eqn:likelihood}  
    L(\boldsymbol{\theta}) = \prod_{i \in \mathcal{U}} (\pi^B_i)^{\delta_i}(1-\pi^B_i)^{(1-\delta_i)}
\end{equation}

and the corresponding log-likelihood function is

\begin{equation} \label{eqn:loglikelihood}   
    l(\boldsymbol{\theta}) = \sum_{i \in \mathcal{U}}\delta_i \text{log}\pi^B_i + \sum_{i \in \mathcal{U}}(1 - \delta_i)\text{log}(1 - \pi^B_i)
\end{equation}

The two components in (\ref{eqn:loglikelihood}) can be estimated in various ways, and in this section we provide an overview of several approaches that have previously been proposed.

\citet{chen.etal_2020} re-arrange the log-likelihood function so that it becomes the sum of two terms: one which is over $B$, and one which is over $\mathcal{U}$.  The reference sample $A$ is used to estimate the term that is over $\mathcal{U}$, obtaining the pseudo-likelihood function

\begin{equation} \label{eqn:clw.loglikelihood}   
    l_{CLW}(\boldsymbol{\theta}) = \sum_{i \in B}\text{log} \Bigl( \frac{\pi^B_i}{1 - \pi^B_i} \Bigr) + \sum_{i \in A}d^A_i \text{log}(1 - \pi^B_i)
\end{equation}

\citet{kim.wang_2019} proposed a data integration approach to propensity score estimation, assuming that it is possible to identify membership in $B$ for each unit in the reference sample $A$, thus obtaining the value of $\delta_i$ for all units in $A$.  The components in (\ref{eqn:loglikelihood}) can then be estimated using the reference sample $A$ alone, resulting in the pseudo-likelihood function

\begin{equation} \label{eqn:KW.loglikelihood}   
    l_{KW}(\boldsymbol{\theta}) = \sum_{i \in A}d^A_i \delta_i \text{log}\pi^B_i + \sum_{i \in A}d^A_i (1 - \delta_i)\text{log}(1 - \pi^B_i)
\end{equation}

\citet{valliant.dever_2011} re-write (\ref{eqn:loglikelihood}) as

\begin{equation} \label{eqn:loglikelihood2}   
    l(\boldsymbol{\theta}) = \sum_{i \in B} \text{log}\pi^B_i + \sum_{i \in \mathcal{U} \setminus B} \text{log}(1 - \pi^B_i)
\end{equation}

and form the pseudo-likelihood function

\begin{equation} \label{eqn:VD.loglikelihood}   
    l_{VD}(\boldsymbol{\theta}) = \sum_{i \in B}\text{log}\pi^B_i + \sum_{i \in A}w^*_i \text{log}(1 - \pi^B_i)
\end{equation}

where $w^*_i=d^A_i(\hat{N} - N_B)/\hat{N}$ are scaled weights for the sample $A$ such that $\sum_{i \in A} w^*_i = \hat{N} - N_B$, and $\hat{N} = \sum_{i \in A}d^A_i$.

\citet{wang.etal_2021} developed the Adjusted Logistic Propensity (ALP) method by first creating an artificial population $\mathcal{U}_c$ consisting of $B$ stacked on to $\mathcal{U}$.  The elements in $B$ appear in $\mathcal{U}_c$ twice, and each copy of an element is considered to be distinct from the other.  The authors define an indicator $r_i, i \in \mathcal{U}_c$ which takes the value 1 if $i \in B$, and 0 if $i \in \mathcal{U}$.  Further, they define $p_i = P(r_i = 1) = P(i \in B|\mathcal{U}_c)$.  The likelihood varies depending on which units are in $B$, and can be expressed as

\begin{equation} \label{eqn:WVL likelihood}
    L^*(\boldsymbol{\theta}) = \prod_{i \in \mathcal{U}_c} (p^{r_i}_i)(1-p_i)^{1-r_i}
\end{equation}

while the corresponding log-likelihood is

\begin{equation} \label{eqn:WVL.loglikelihood}   
    l^*_{WVL}(\boldsymbol{\theta}) = \sum_{i \in \mathcal{U}}\delta_i\text{log}p_i + \sum_{i \in \mathcal{U}} \text{log}(1 - p_i)
\end{equation}

The second term in (\ref{eqn:WVL.loglikelihood}) will need to be estimated when the unit level information for $\mathcal{U}$ is not known.  Using $A$ to estimate the second term, the pseudo-likelihood function obtained is

\begin{equation} \label{eqn:WVL.ploglikelihood}   
    l_{WVL}(\boldsymbol{\theta}) = \sum_{i \in B}\text{log}p_i + \sum_{i \in A} d^A_i \text{log}(1 - p_i)
\end{equation}

For each of the pseudo-likelihood functions described above, the maximum pseudo-likelihood estimators $\hat{\boldsymbol{\theta}}$ can be produced by deriving the corresponding score function, setting it to zero and then solving for $\boldsymbol{\theta}$.  Generally, a numerical procedure such as Newton-Raphson will need to be used to obtain a solution.

For the ALP method, once we have obtained $\hat{\boldsymbol{\theta}}$ by maximising (\ref{eqn:WVL.ploglikelihood}), we can obtain $\hat{\pi}^B_i$ through the relationship

\begin{align}
    \hat{\pi}^B_i = \frac{\hat{p}_i}{1 - \hat{p}_i}
\end{align}

\section{Overlap-Excluded Propensity Score Estimation}\label{oe_propscores}

Like \citet{kim.wang_2019}, we assume that it is possible to identify membership in $B$ for each unit in the reference sample $A$.  Note that the first component in (\ref{eqn:loglikelihood2}) can be calculated directly using $B$.  Our main idea is to have the second component estimated using $A \setminus B$, that is, by the sample in $A$ that are not also in $B$.  The resulting pseudo-likelihood is given by

\begin{equation} \label{eqn:new.loglikelihood}   
    l_{OE}(\boldsymbol{\theta}) = \sum_{i \in B}\text{log}\pi^B_i + \sum_{i \in A \setminus B}d^A_i \text{log}(1 - \pi^B_i)
\end{equation}

We term this the Overlap-Excluded (OE) pseudo-likelihood function.  Note that $l_{OE}$ is very similar to $l_{WVL}$, with the differences being that our proposed approach i) directly models $\pi^B_i$ rather than the indirect probability $p_i$, and ii) the second sum is over $A \setminus B$ rather than over all $A$.

Under a logistic regression model for the propensity scores, $\pi^B_i = \text{exp}(\boldsymbol{x}^T_i\boldsymbol{\theta}_0)/(1 + \text{exp}(\boldsymbol{x}^T_i\boldsymbol{\theta}_0))$.  Then the pseudo-likelihood function (\ref{eqn:new.loglikelihood}) becomes

\begin{align} \label{eqn:new.pseudo.ll}
    l_{OE}(\boldsymbol{\theta}) 
    &= \sum_{i \in B} \Bigl\{ \boldsymbol{x}^T_i\theta - \text{log}(1 + \text{exp}(\boldsymbol{x}^T_i\theta)) \Bigr\} - \sum_{i \in A \setminus B} d^A_i \text{log}(1 + \text{exp}(\boldsymbol{x}^T_i\theta))
\end{align}

The score function $S(\boldsymbol{\theta}) = \frac{\partial }{\partial \boldsymbol{\theta}}l_{OE}(\boldsymbol{\theta})$ is given by

\begin{align} \label{eqn:score}
    S(\boldsymbol{\theta})
    &= \sum_{i \in B} \boldsymbol{x}_i - \sum_{i \in B}\pi(\boldsymbol{x}_i,\boldsymbol{\theta})\boldsymbol{x}_i - \sum_{i \in A \setminus B}d^A_i\pi(\boldsymbol{x}_i,\boldsymbol{\theta})\boldsymbol{x}_i
\end{align}

We can obtain the maximum pseudo-likelihood estimator $\hat{\boldsymbol{\theta}}$ by solving $S(\boldsymbol{\theta}) = \boldsymbol{0}$.  A solution can be found by applying the Newton-Raphson iterative procedure

\begin{align} \label{eqn:new.hessian}
    \boldsymbol{\theta}^{(m+1)} &=  \boldsymbol{\theta}^{(m)} + \Bigl\{ H(\boldsymbol{\theta}^{(m)}) \Bigr\}^{-1} S(\boldsymbol{\theta}^{(m)}) \nonumber \\
    \text{where} \nonumber \\
    H(\boldsymbol{\theta}^{(m)}) &= - \frac{\partial^2 }{\partial \boldsymbol{\theta}^2} l_{OE}(\boldsymbol{\theta}) \nonumber \\
    &=  \sum_{i \in B} \pi(\boldsymbol{x}_i,\boldsymbol{\theta})(1 - \pi(\boldsymbol{x}_i,\boldsymbol{\theta}))\boldsymbol{x}_i\boldsymbol{x}^T_i + \sum_{i \in A \setminus B}d^A_i\pi(\boldsymbol{x}_i,\boldsymbol{\theta})(1 - \pi(\boldsymbol{x}_i,\boldsymbol{\theta}))\boldsymbol{x}_i\boldsymbol{x}^T_i
\end{align}

The OE propensity score for unit $i$ is obtained from the maximum pseudo-likelihood estimates of $\hat{\boldsymbol{\theta}}$ through

\begin{align}
    \hat{\pi}^B_i = \frac{\text{exp}(\boldsymbol{x}^T_i\boldsymbol{\hat{\theta}})}{(1 + \text{exp}(\boldsymbol{x}^T_i\boldsymbol{\hat{\theta}}))}
\end{align}

\subsection{Inverse Probability Weighted Estimator}

The estimated propensity scores $\hat{\pi}^B_i$ from Section \ref{oe_propscores} may be used to produce an Inverse Probability Weighted (IPW) estimator (\citealt{chen.etal_2020}, \citealt{wu_2022a}) of the population mean:

\begin{equation} \label{eqn:IPW}
    \hat{\mu}_{IPW} = \frac{1}{\hat{N}_B}\sum_{i \in B}\frac{y_i}{\hat{\pi}^B_i}
\end{equation}

where $\hat{N}_B = \sum_{i \in B}(\hat{\pi}^B_i)^{-1}$.  The propensities may also be used in a doubly robust manner such as described in \citet{chen.etal_2020}.

\subsection{Plug-in Variances}

The variance of the IPW estimator (\ref{eqn:IPW}) can be estimated using the plug-in estimator given by
        
\begin{align}
    \hat{Var}(\hat{\mu}_{IPW})
    &= \frac{1}{N^2}\sum_{i \in B} (1-\hat{\pi}^B_i) \Bigl[\frac{(y_i - \hat{\mu}_{IPW})}{\hat{\pi}^B_i} - \boldsymbol{\hat{b}}^T\boldsymbol{x}_i  \Bigr]^2 + \label{eqn:varIPW CLW term} \\
    & \boldsymbol{\hat{b}}^T\boldsymbol{\hat{V}}\boldsymbol{\hat{b}} \text{ } +  \label{eqn:varIPW desvar} \\
    &  \frac{1}{N^2}\sum_{i \in B}\Bigl[(\hat{\pi}^B_i)^2(1 - \hat{\pi}^B_i)(d^A_i - 1) \boldsymbol{\hat{b}}^T\boldsymbol{x}_i\boldsymbol{x}^T_i\boldsymbol{\hat{b}} \Bigr] \label{eqn:varIPW addterm}
\end{align}

where $\boldsymbol{\hat{b}}^T = \sum_{i \in B}(\frac{1}{\hat{\pi}^B_i} - 1)(y_i - \hat{\mu}_{IPW})\boldsymbol{x}^T_i \Bigl[ \sum_{i \in B}(1 - \hat{\pi}^B_i)\boldsymbol{x}_i\boldsymbol{x}^T_i \Bigr]^{-1}$, $\hat{\pi}^B_i$ is the estimated propensity for unit $i$ obtained using $\hat{\boldsymbol{\theta}}$, and $\boldsymbol{\hat{V}} = \frac{1}{N^2} \hat{Var}_d \Bigl(\sum_{i \in A}d^A_i\hat{\pi}^B_i(1-\hat{\pi}^B_i)\boldsymbol{x}_i \Bigr)$, where $\hat{Var}_d(\cdot)$ denotes the design-based variance estimator.  Note that \citet{chen.etal_2020} use a slightly different plug-in estimator for $\boldsymbol{\hat{b}}^T$ which would be more appropriate if $B$ is small relative to $A$.

Component (\ref{eqn:varIPW addterm}) requires knowledge of the reference sample design weights for units in $B$, and this may not be obtainable.  An alternative plug-in estimator for (\ref{eqn:varIPW addterm}) which does not require knowledge of the design weights for units in $B$ is
\begin{align}
    \frac{1}{N^2}\sum_{i \in A} (d^A_i)^2\Bigl[(\hat{\pi}^B_i)^3(1 - \hat{\pi}^B_i)
    \hat{\boldsymbol{b}}^T\boldsymbol{x}_i\boldsymbol{x}^T_i\hat{\boldsymbol{b}} \Bigr] - 
    \frac{1}{N^2}\sum_{i \in B} \Bigl[(\hat{\pi}^B_i)^2(1 - \hat{\pi}^B_i)\hat{\boldsymbol{b}}^T\boldsymbol{x}_i\boldsymbol{x}^T_i\hat{\boldsymbol{b}} \Bigr]
\end{align}

\section{Simulation Study} \label{simstudy}

A simulation study was conducted to examine the performance of the IPW estimates formed using the OE propensity scores.  The setup of the simulation is similar to what is used in \citet{chen.etal_2020}.  In our simulation we consider a finite population of size $N = 200,000$ that includes a data item of interest $y$ and auxiliary variables $\boldsymbol{x}$, related by the model

\begin{equation}
    y_i = f(\boldsymbol{x},\boldsymbol{\beta}) = 2 + x_{1i} + x_{2i} + x_{3i} + x_{4i} + \sigma\epsilon_i \nonumber
\end{equation}

with $x_{1i} = z_{1i}, x_{2i} = z_{2i} + 0.3x_{1i}, x_{3i} = z_{3i} + 0.2(x_{1i} + x_{2i}), x_{4i} = z_{4i} + 0.1(x_{1i} + x_{2i} + x_{3i})$, and $z_{1i} \sim \text{Bernoulli}(0.5), z_{2i} \sim \text{Uniform}(0.2), z_{3i} \sim \text{Exponential}(1)$, and $z_{4i} \sim \chi^2(4)$.  The error terms $\epsilon_i$ are independent and identically distributed as $N(0,1)$.  In our simulations, we utilise two values of $\sigma$, chosen such that the correlation coefficient $\rho$ between $y$ and $\boldsymbol{x}^T\boldsymbol{\beta}$ is either 0.3 (moderately weak) or 0.7 (moderately strong).

The true propensity scores $\pi^B_i$ are given by

\begin{equation}
    \text{log}\Bigl(\frac{\pi^B_i}{1 - \pi^B_i} \Bigr) = \theta_0 + 0.1x_{1i} + 0.2x_{2i} + 0.3x_{3i} + 0.4x_{4i} \nonumber
\end{equation}

where the intercept $\theta_0$ is chosen to ensure the target sample size for $N_B$ is achieved.  Poisson sampling is used to select the sample $B$, while the randomised systematic probability proportional to size sampling (PPS) method is used for the probability sample $A$.  In general, the inclusion probabilities for $\pi^A_i$ are proportional to $z_i = c + x_{3i}$, where $c$ is chosen to ensure the ratio max($z_i$)/min($z_i$) = 10.  Different sets of simulations were run for three values of $N_B$ - 2,000, 50,000 and 140,000.  The sample size for $A$ was kept fixed at $n_A = 5,000$ for all simulation runs.

The simulation study consisted of the following broad steps:
\begin{enumerate}
    \item Generating the finite population, including the variable of interest $y$ and the auxiliary items $\boldsymbol{x}$
    \item Drawing a random subsample of size $N_B$ from the population to be the ``big dataset''
    \item Drawing a reference probability sample of size $n_A=5,000$
    \item Producing an estimate of the population mean of $y$ using a number of different estimators
\end{enumerate}

Step 1 was conducted once only, while Steps (2) to (4) were repeated $R = 2,000$ times.  The performance of the different estimators was evaluated by calculating the Monte Carlo percentage Relative Bias (\%RB) and percentage Relative Root Mean Squared Error (\%RRMSE):

\[ \text{\%RB} = \frac{1}{R}\sum_{r=1}^{R}\frac{\hat{\bar{Y}}_r - \bar{Y}}{\bar{Y}} \times 100 \]
\[\text{\%RRMSE}=\frac{\sqrt{\frac{1}{R}\sum_{r=1}^{R}(\hat{\bar{Y}}_r-\bar{Y})^2}}{\bar{Y}} \times 100 \]

where $\hat{\bar{Y}}_r$ is the estimate of the population mean computed for the $r$'th simulation run and $\bar{Y}=\mu_Y$ is the true population mean.

Results were produced for the following estimators:

\begin{itemize}
    \item Naive - Simple sample mean from $B$
    \item OE - IPW estimates using the proposed Overlap-Excluded propensity score estimator
    \item CLW - The H\'ajek-like IPW estimator in \citet{chen.etal_2020}
    \item WVL - The ALP estimator proposed by \citet{wang.etal_2021}
    \item KW - The estimator proposed by \citet{kim.wang_2019}
    \item VD - The estimator from \citet{valliant.dever_2011}
\end{itemize}

\subsection{Simulation Results}

Table \ref{tab:sim1_results} provides the \%RB and \%RRMSE values for each of the estimators under the different sample sizes for $B$ and correlations between $y$ and $\boldsymbol{x}^T\beta$.  The Naive estimate based solely on the sample mean of $B$ is biased for all $N_B$ sizes, as expected given the level of undercoverage in $B$.  The VD estimator is also slightly biased for all $N_B$ sizes, which was a finding of \citet{chen.etal_2020}.  The proposed OE approach, along with the CLW, KW and WVL estimators, are all unbiased.

\begin{table}[ht]
    \centering
    \caption{Comparative performance of estimators}
      \begin{tabular}{lrrrrrr}
        \hline
        \hline
      \multicolumn{1}{c}{} & \multicolumn{2}{c}{$N_B=2,000$}
        & \multicolumn{2}{c}{$N_B=50,000$} & \multicolumn{2}{c}{$N_B=140,000$} \\
       & \%RB & \%RRMSE & \%RB & \%RRMSE & \%RB & \%RRMSE \\
        \hline
      \multicolumn{2}{l}{$\rho(y,\boldsymbol{x}^T\beta)=0.3$} &       &       &             &  \\
      Naive & 73.49 & 73.54 & 27.55 & 27.55 & 8.11  & 8.12 \\
      OE    & -0.21 & 5.78  & -0.02 & 0.82  & 0.01  & 0.35 \\
      CLW   & -0.31 & 6.17  & -0.12 & 1.34  & -0.06 & 0.82 \\
      WVL   & 0.93  & 5.47  & 0.44  & 0.91  & 0.04  & 0.59 \\
      KW    & -0.21 & 7.20  & -0.04 & 1.20  & -0.01 & 0.47 \\
      VD    & 4.80  & 7.21  & 9.80  & 9.83  & 5.79  & 5.79 \\
      \multicolumn{2}{l}{$\rho(y,\boldsymbol{x}^T\beta)=0. 7$} &       &       &            &       &  \\
      Naive & 73.73 & 73.74 & 27.62 & 27.62 & 8.14  & 8.14 \\
      OE    & -0.20 & 2.71  & -0.03 & 0.55  & 0.01  & 0.30 \\
      CLW   & -0.30 & 3.40  & -0.13 & 1.21  & -0.06 & 0.80 \\
      WVL   & 0.95  & 2.39  & 0.44  & 0.72  & 0.04  & 0.56 \\
      KW    & -0.25 & 4.95  & -0.05 & 1.06  & -0.01 & 0.44 \\
      VD    & 4.83  & 5.57  & 9.82  & 9.83  & 5.80  & 5.81 \\
      \hline
      \end{tabular}
    \label{tab:sim1_results}
  \end{table}

In terms of the RRMSE performance of the estimators, our proposed OE approach provided the lowest RRMSEs when $N_B$ was 50,000 or 140,000, while the WVL estimator provided the most efficient estimates for the scenarios when $N_B$ was small.  The KW estimator did not perform well compared with most of the other estimators when $N_B$ was small, however its results improved under larger values of $N_B$.  For the scenario with $N_B=140,000$, the KW estimator resulted in the second lowest RRMSE values, after the OE approach.

For the smallest value of $N_B$, the VD estimator produced the third lowest RRMSE values, after the WVL and the OE approaches.  Under this scenario, we expected that there would likely only be a minimal overlap between the samples $A$ and $B$.  Nevertheless, the RB for the estimator was still non-negligible.  Under larger values of $N_B$ the RB of the VD estimator becomes the dominant contributor to the overall RRMSE value and its performance deteriorated relative to other estimators.

\begin{table}[ht]
    \centering
    \caption{95\% Coverage Probability}
      \begin{tabular}{lrr}
        \hline
        \hline
        \\[-5pt]
      \multicolumn{1}{c}{} & \multicolumn{2}{c}{$\rho(y,\boldsymbol{x}^T\beta)$} \\
      \multicolumn{1}{c}{$N_B$} & \multicolumn{1}{c}{0.3}   & \multicolumn{1}{c}{0.7} \\
        \hline
        2,000  & 0.946 & 0.950 \\
        50,000 & 0.953 & 0.951 \\
        140,000 & 0.951 & 0.948 \\
        \hline
      \end{tabular}
    \label{tab:coverage_rate}
  \end{table}

Table \ref{tab:coverage_rate} provides coverage probabilities for the 95\% confidence interval using the plug-in variance estimator for the proposed OE approach.  The coverage probabilities are all close to the expected value of 0.95.

\section{Concluding Remarks} \label{conclusion}

Quasi-randomisation approaches have been developed to enable finite population inferences to be formed from non-probability data, including the big datasets that are the focus of this paper.  These approaches utilise available auxiliary information from the population or from a reference probability sample to estimate a propensity of selection into the big dataset.  In this paper, we have proposed a new approach for estimating these propensities using what we have termed the Overlap-Excluded (OE) approach to estimating propensity scores.  This estimator takes advantage of the ability to integrate the reference sample with the big dataset.

When the model for $\hat{\pi}^B_i$ is correctly specified, the IPW estimator formed from the OE propensity scores is unbiased.  The estimates are also efficient relative to the other approaches we have considered in this paper, particularly when the big data sample is large.  While not included in this paper, it can be shown theoretically (confirming the simulation results in this paper) that when the reference sample is taken by Poisson sampling the OE IPW estimator is at least as efficient as the H\'ajek-like estimator from \citet{chen.etal_2020} and the \citet{kim.wang_2019} data integration estimator.  Further, it is more efficient than the ALP estimator from \citet{wang.etal_2021} when $\pi^B_i$ is large.

When applying the propensity score estimation approaches considered in this paper, the reference sample is generally taken as a given.  In other words, there exists a big dataset that we would like to use to make finite population inferences, and we find an existing probability sample that can be used in the propensity score estimation.  However, we may instead wish to design an efficient reference sample that can then be used to supplement the big dataset in order to achieve a particular accuracy target.  Such judicious sample design using optimal selection probabilities can lead to reduced respondent burden and cost savings for the survey organisation.  This will be a future research focus for the authors.

\section*{Acknowledgements and Disclaimer}

The first author was supported by funding from the Australian Bureau of Statistics and the Sir Roland Wilson Foundation.

The views expressed in this paper are those of the authors and do not necessarily represent the views of the Australian Bureau of Statistics.  Where quoted or used, they should be attributed clearly to the authors.


\bibliography{./Journal_Paper/PSEstimator.bib}

\end{document}